# Estimating the infection horizon of COVID-19 in eight countries with a data-driven approach

G. D. Barmparis[1] and G. P. Tsironis[2]

Department of Physics, University of Crete, Heraklion 71003, Greece

[1]barmparis@physics.uoc.gr, [2]gts@physics.uoc.gr

10 April 2020

Abstract

The COVID-19 pandemic has affected all countries of the world producing a substantial number of fatalities accompanied by a major disruption in their social, financial and educational organization. The strict disciplinary measures implemented by China were very effective and thus were subsequently adopted by most world countries to various degrees. The infection duration and number of infected persons are of critical importance for the battle against the pandemic. We use the quantitative landscape of the disease spreading in China as a benchmark and utilize infection data from eight countries to estimate the complete evolution of the infection in each of these countries. The analysis predicts successfully both the expected number of daily infections per country and, perhaps more importantly, the duration of the epidemic in each country. Our quantitative approach is based on a Gaussian spreading hypothesis that is shown to arise as a result of imposed measures in a simple dynamical infection model. This may have consequences and shed light in the efficiency of policies once the phenomenon is over.

## 1. Introduction

The epidemic spread of the new coronavirus (SARS-CoV-2 or COVID-19) first in the Wuhan province of China and subsequently in the rest of the world has generated a substantial response from states in an effort to contain the spreading and eventually eliminate the threat [1]. The COVID-19 has induced substantial number of deaths in all continents, although it appears that Europe and the US in North America are the hardest hit areas so far. The response of the medical profession to the challenge was fast and heroic, but a large part of the responsibility for containment lies in the policy makers who are forced to take critical decisions with limited factual information. Predictions about the immediate future spreading of the epidemic and the resulting fatalities vary wildly depending critically on models used and parameters estimated though the models. Thus, it would be helpful for the immediate understanding of the epidemic evolution to resort, to the extent possible, to data driven estimates [2,3]. The present analysis is based on the fact that we have already significant, quantitative knowledge on the COVID-19 spread dynamics and this comes from China. In China the virus appeared on December 23, 2019 in the Wuhan region and after its fast-initial spreading, strict rules of social distancing were imposed almost a month later. It appears that now, approximately three months after the initially reported cases, the spreading in China has subsided. It is noteworthy, that most countries in the world eventually applied strict or milder rules for social distancing in the spirit of China. It is thus fair to say, that most countries, in one way or the other, expect and definitely hope to follow the China pattern with a minimal loss of life.

## 2. Data analysis and projections

The present work uses both the knowledge from the China experience as well as the hard infection data from each country. We use China in the sense that we analyze the distribution of the daily reported numbers of infected persons as a function of time and find that, with the exception of some outliers, it follows a Gaussian function. Armed with this knowledge and given the fact that most countries follow more or less the Chinese approach of social distancing, we assume the evolution will be similarly qualitatively Gaussian, although clearly with differences. The latter will be reflected in three parameters of the individual country distribution, viz. its mean, standard deviation and peak value; we thus surmise that these three numbers specify the details of the virus spreading in each country. Based on this assumption we fit through a simulated annealing-like process available country data and obtain a full predicted evolution of the spreading in each country. Since the China-derived Gaussian evolution is critical for this work, we employ additionally a standard SIR infection model and derive a similar dependence. The simple predictive method we use in this work has appealing features: it is easy to implement, uses specific prior knowledge, i.e. that of China and the SIR model and being data driven it provides specificity.

We select eight counties, viz. Greece, the Netherlands, Germany, Italy, Spain, France, the UK and the USA and use data reported on April 4, 2020. In Table 1, we show a comparison between reported and predicted numbers of infected individuals; we note that the relative error is reasonably small for most countries. In Fig. 1, we show the quantitative predictions of this analysis for the eight selected countries; we show three predicted curves based on either all points prediction (using all days, red dashed line), without the last point predictions (using all days except the last one, green dashed dotted line) or without the last two points predictions (using all days except the last two, black dotted line) as well as the available data (blue points and line). These three curves demonstrate to some extent the degree of uncertainty of the predicted values and horizon. We observe that an effective curve flattening has occurred in Greece while both the UK and the USA seem to be on the infection rise with the latter to be approaching a sharp maximum. It is noteworthy that based on this analysis Italy, Spain and the Netherlands seem to have passed the highest point of infection while Greece, France, and Germany are about to pass it as well.

| Country | Total cases reported on April 4 | | Error $\frac{|\Delta x|}{x}$ (%) | Predicted | | |
|---|---|---|---|---|---|---|
| | Reported [4] | Predicted | | Peak date | Horizon Date (4σ) | Total # of cases |
| Greece | 1673 | 1621 | 3.0 | 04/03 | 05/18 | 2811 |
| Netherlands | 16627 | 16862 | 1.4 | 03/31 | 05/05 | 23713 |
| Germany | 91622 | 90460 | 1.3 | 04/02 | 05/08 | 140003 |
| Italy | 124632 | 129180 | 3.6 | 03/26 | 05/08 | 156975 |
| Spain | 124736 | 129628 | 3.9 | 03/31 | 05/02 | 173535 |
| France | 68605 | 69330 | 1.1 | 04/05 | 05/21 | 141973 |
| UK | 41903 | 42888 | 2.4 | 04/12 | 05/26 | 165443 |
| USA | 312237 | 315677 | 1.1 | 04/05 | 05/10 | 654207 |

Table 1. Total number of infections reported on April 4, 2020 and the corresponding predictions obtained from our model. In the second part of the table we give the predicted dates for the peak of the infection, its horizon (the date at 4σ of the distribution after the peak) and the total predicted number of reported infections.

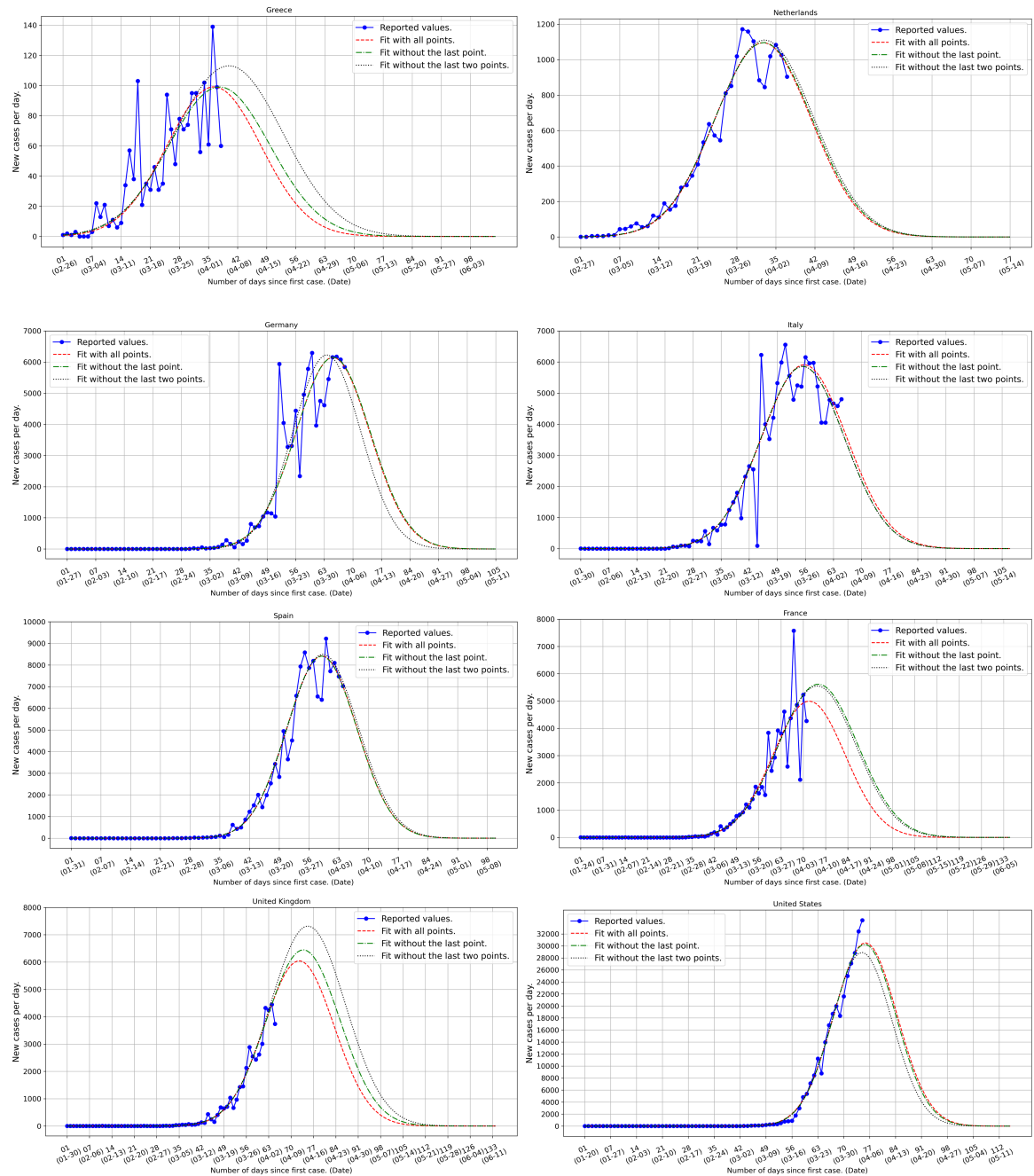

Figure 1. Country level estimates of daily number of infections based on the available country data (blue points and line) reported on April 4, 2020 [4]. The red dashed lines give the predicted evolution of the infection based on all available data up to and including the ones of the last reported day. The green dashed dotted lines include the data up to a day earlier than the last reported date while the black dotted lines include the data up to two days earlier than the last reported date. The difference in the three predicted curves, red, blue and black, reflects thus the relative robustness of the phenomenon and gives an estimate of the fluctuations. A more complete statistical analysis of the infection horizon will be presented as more data accumulate [5]. From the figures we see that, for instance, in the case of Greece where strict rules were imposed early, both the number of infections and the "flattening of the curve" is occurring in a rather controlled way while, on the contrary, in Spain the peak is more sharp and with a vastly larger number of infections.

In Fig. 2, we present the mean predicted date and its standard deviation over the last 5 daily runs of the model (since March 31, 2020) for the peak and the horizon date for each country in this work. Countries like Greece, France, Germany, the UK and the USA that have not passed the peak yet have a large dispersion (about 2 weeks) of the predicted peak and horizon dates while countries like Italy, Spain, and the Netherlands that seems to have passed the peak date give more robust predictions.

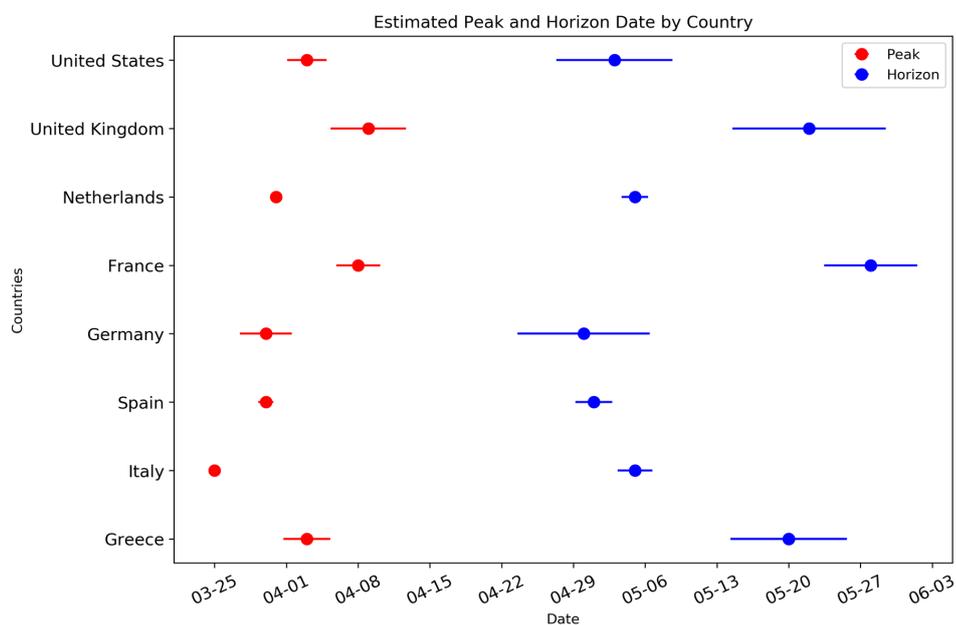

Figure 2. Peak (red) and Horizon (blue) date mean value and standard deviation for each country considered in this work.

## 3. China data analysis

We turn now to China that provides the data basis for this analysis. In China 82295 cases have been reported by March 31. If we ignore the extreme event of reporting 15141 infected cases in February 13, 2020 China has 67154 cases in total. Our fit for China predicts 70306 total cases or 66955 excluding the prediction for the same day; this results to an error of 0.3% (or less than 200 infected cases). We note that the symmetry of the Gaussian function used in this work for the prediction of the evolution of the pandemic is dictated directly by the Chinese data.

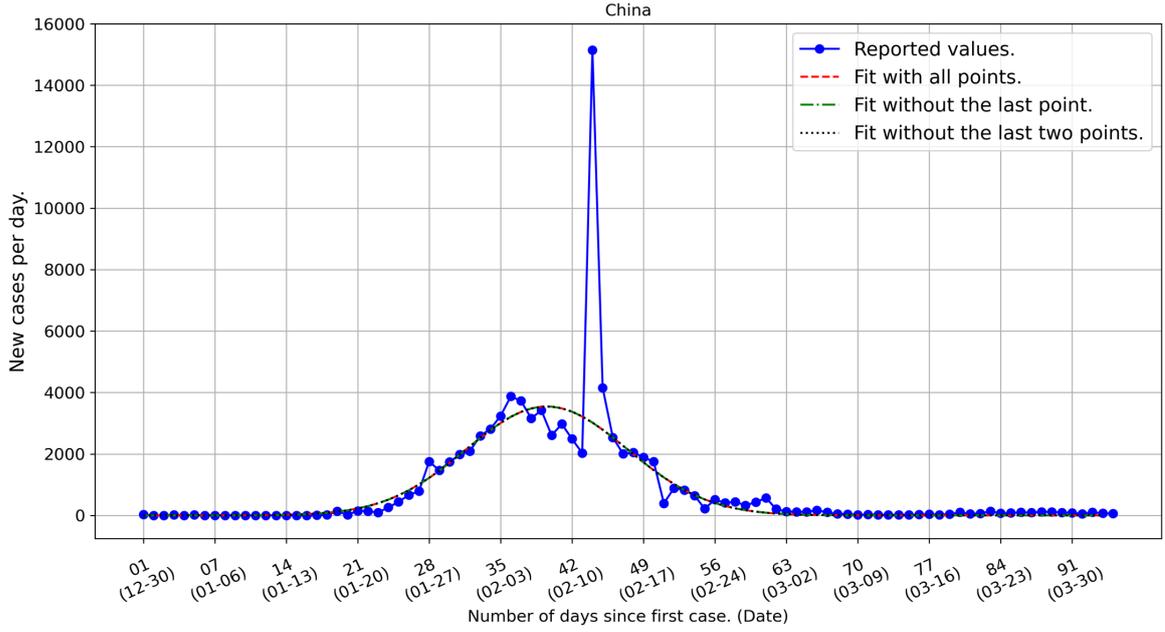

Figure 3. COVID-19 infected individuals in China during the period 31 December 2019 to 31 March 2020. We note the large outlier on February 13, 2020 related to reporting issues. If we exclude this singular event the assumed compete circle of infection in China follows a Gaussian function with mean, standard deviation and height equal to 40.5 days, 7.9 days and 3557 cases, respectively. The infection horizon that could be defined at 4σ of the distribution is approximately equal to 2 months from the onset of the infection.

## 4. Model-based justification of the Gaussian fitting Hypothesis

The critical assumption for the projection of the evolution of the infected persons is that of the Gaussian evolution. Although this trend is data-driven we show here that it may be justified in the context of the standard SIR model. In the latter, in a fixed population of individuals we denote with S, I and R, the Susceptible, Infected and Recovered or Removed percentages of persons involved in the infection. Clearly $S + I + R = 1$. The model is defined with two equations for the susceptible and infected individuals since the third equation follows from the sum constraint. We have

$$\frac{dS}{dt} = -aSI \qquad (1)$$

$$\frac{dI}{dt} = aSI - \mu I \qquad (2)$$

where $\alpha, \mu$ are the infection and recovery rates respectively. In Eq. (1) the time derivative is always negative and thus the susceptible population always decreases. On the other hand, from Eq. (2) the condition for flattening of the infection growth where the derivative is zero happens at a critical susceptible number, i.e. for $S_c = \mu/\alpha$. While $S > S_c$ the infected population grows, reaches a maximum at $S = S_c$ and subsequently decays to zero and the infection ends with all individuals either recovered or actually removed from the population.

The values of the two parameters $\alpha$ and $\mu$ are critical for the evolution of the infection. The value of $\alpha$ determines how infectious is the spreading; large values infect large population and only at small number of susceptible the infection decays. The value of $\mu$ on the other hand controls the rate at which the individuals do not participate any more in the infection process; large values of $\mu$ result in a very fast decay of the infection. A typical evolution is shown in Fig. 4a where the infected population is seen to grow fast reach a maximum and subsequently have a relatively slow decay. The time evolution is distinctly non-Gaussian.

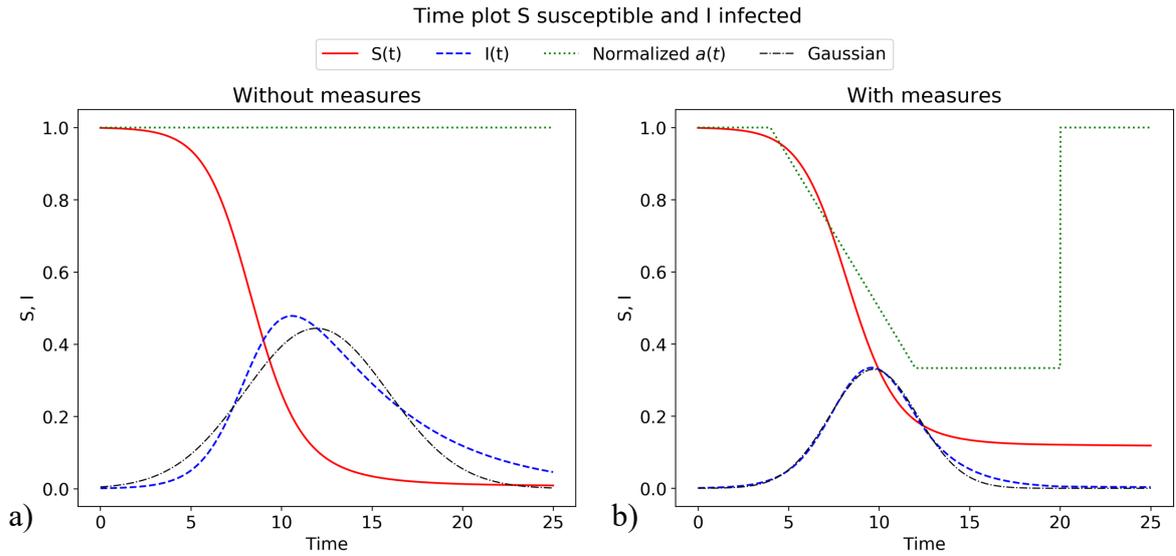

Figure 4. Time evolution of susceptible S(t) (red solid line), infected I(t) (blue dashed line) percentages and the normalized infection rate (green dotted line) with the corresponding Gaussian approximation for the I(t) (black dashed dotted line). a) No additional measures (constant infection rate) and Gaussian parameters, height = 0.445, mean = 11.92, standard deviation = 3.94, PCC = 0.964, and b) with measures (time-dependent infection rate) and Gaussian parameters, height = 0.330, mean = 9.70, standard deviation = 2.44, PCC = 0.998.

We now assume that specific measures are taken in the processes of the infection; this can be easily implemented in the SIR model by taking the infection rate to be time-dependent, i.e. $\alpha = \alpha(t)$. When, for instance in the model we start with a given value of the infection parameter and measures of social distancing are introduced, the value of the infection parameter is reduced. While a more general analysis can be done easily [5], we focus here in one specific case that is relevant to the form of application of measures for the COVID-19 pandemic. In Fig. 4b, we show the time evolution for the case where measures of social distancing where applied gradually and kept until the end of the pandemic. We see that the actual time evolution of the infected population is not only distinctly Gaussian but, more importantly, there is not even a damped recurrency of the infection. It is noteworthy that this rather optimistic scenario of measure imposition that gradually decrease social distancing and thus the infection rate, is what approximately happened in most countries. We quantify the correlation between the infection curve and the fitted Gaussian function using the Pearson Correlation Coefficient (PCC) that measures the statistical relationship between two curves (PCC = -1 means perfectly anti-correlated curves, PCC = 0 non-correlated and PCC = 1 perfectly correlated curves). In the case without measures the PCC between the curve of the infection and the fitted Gaussian equals 0.964, while with measures is equal to 0.998 indicating an almost perfect match between the two curves.

Finally, a simple mathematical indication for the Gaussian-like in time behavior of the infection evolution can be obtained from an approximate analysis of Eqs. (1), (2). We note that the imposition of gradual social distancing through a linear drop in the infection rate leads, to lowest order, to a linear drop in the susceptible population. Assuming thus a dependence of the form $S(t) = \gamma - \beta t$, where $\beta, \gamma$ appropriate constants (to lowest order) and substituting this form to Eq. (2) we obtain a solution of the latter that is basically Gaussian, i.e

$$I(t) = I_c e^{-\frac{\alpha\beta}{2}t^2 + (\alpha\gamma - \mu)t} \qquad (3)$$

where $I_c$ the initial infection rate at time negatively large for the present form of the solution. We can argue that the additional exponential dependence in the approximate solution of Eq. (3) is very weak leading to the Gaussian exponential form found also through the numerical analysis and shown in Fig. 4b.

## 5. Conclusions

The analysis presented is based on the historical behavior of the COVID-19 spreading in China with its resulting Gaussian evolution hypothesis as well as the available data of infected persons in eight countries. It gives an estimate of the infection horizon in each country as well as the expected number of infected persons. While in countries like Greece that imposed early strict rules of social distancing the infected numbers are relatively low, in other countries such as the UK or the USA we see much larger numbers and different recovery horizons. Further analysis with a complete list of countries will be presented elsewhere [5]. We find empirically that the China virus infection follows a Gaussian in time evolution. Although this feature appears initially to be at odds with the simple SIR model, it nevertheless follows from it when gradual social distancing measures are imposed. Furthermore, provided the gradual measure imposition is retained, the mathematical model does not predict an infection recurrence. The difference in the behavior among countries based on the available infection data is reflected in the values of the fitted parameters for the individual countries' distributions. While the data-driven model appears to work reasonably well with most countries, it seems that in the present phase of the infection it is not as close to the peak dynamics in the US and perhaps UK. This might be related to the different approaches in social distancing measures taken by both countries. The ultimate success or failure of this model will be judged in due time when relevant conclusions can be drawn with more certainty. We hope that this work gives just an upper limit in the behavior of the COVID-19 pandemic since other factors such as ambient temperature rise, increase in the available medical support and change of human behavior will hopefully assist in the faster containment of the spreading. We believe that once the pandemic is over useful policy conclusions can be drawn from the successes and failures of the present model.

**Appendix: Computational method**

For each one of the selected countries (eight countries plus China that is the benchmark country) we approximate the number of new cases per day, *NCPD(x)*, with a Gaussian function,

$$NCPD(x) = ae^{\frac{-(x-\mu)^2}{2\sigma^2}} \qquad (A1)$$

where *x*, counts the number of days since the first case in that country (day one is the day of the first reported infected case in each country), and the three fitting parameters α, μ and σ determine the height, the position of the peak and the width of the Gaussian, respectively. We initialize each fitting parameters with a randomly assigned value within a reasonable range of values. Subsequently, we use simulated annealing (SA) [6] to find the global minimum of the mean absolute error (MAE) between the reported values (RV) [4] and the predicted ones by Eq. (A1):

$$MAE = \frac{1}{N}\sum_{x=1}^{N} |RV(x) - NCPD(x)| \qquad (A2)$$

In Eq. (A2), *N* is the number of days from the first case in each country until "today", i.e. till 4 April 2020. The minimization process is performed iteratively; a step where MAE reduces compared to the previous step value is accepted and its relevant α, μ and σ parameter values registered. Subsequently, Gaussian random numbers with these new parameters are used as simulated infected data, compared with the available infected country data and the process of MAE stochastic minimization is repeated. The iteration stops when the parameters converge to the optimal ones.